\documentclass{article}

\usepackage[utf8]{inputenc} 
\usepackage[T1]{fontenc}    
\usepackage{hyperref}       
\usepackage{url}            
\usepackage{booktabs}       
\usepackage{amsfonts}       
\usepackage{nicefrac}       
\usepackage{microtype}      
\usepackage{color}
\usepackage{gensymb}
\usepackage{graphicx}
\usepackage{subcaption}
\usepackage{enumerate}
\usepackage{amsmath}
\usepackage{amssymb}

\usepackage{algorithm}  
\usepackage{algorithmicx}  
\usepackage{algpseudocode}  
\usepackage{amsmath}  

\title{Early Detection of Retinopathy of Prematurity (ROP) in Retinal Fundus Images Via Convolutional Neural Networks}

\usepackage[T1]{fontenc}
\usepackage{authblk}
\author[1,2]{Xin Guo}
\author[2]{Yusuke Kikuchi}
\author[1]{Guan Wang\thanks{\emph{Corresponding author. } E-mail: wangguan17@mails.tsinghua.edu.cn} }
\author[3]{Jinglin Yi}
\author[3]{Qiong Zou}
\author[3]{Rui Zhou}
\affil[1]{Tsinghua-Berkeley Shenzhen Institute, China}
\affil[2]{University of California, Berkeley, USA}
\affil[3]{Affiliated Eye Hospital of Nanchang University, China}
\begin{document}

\maketitle

\begin{abstract}
Retinopathy of prematurity (ROP) is an abnormal blood vessel development in the retina of  a prematurely-born infant or an infant with low birth weight. 
ROP is one of the leading causes for infant blindness globally. Early detection of ROP is critical to slow down and avert the progression to vision impairment caused by ROP.
Yet there is limited awareness of ROP even among medical professionals. Consequently, dataset for ROP is limited if ever available, and is in general extremely imbalanced in terms of the ratio between negative images and positive ones. 

In this study, we formulate the  problem of detecting ROP in  retinal fundus images in an optimization framework, and apply  state-of-art convolutional neural network techniques to solve this problem. Experimental results based on our models achieve 100 percent sensitivity, 96 percent specificity, 98 percent accuracy, and 96 percent precision.  In addition, our study shows that as the network gets deeper, more significant features can be  extracted for better understanding of ROP. 

\end{abstract}

\section{Introduction}
Retinopathy of prematurity (ROP) is an abnormal blood vessel development in the retina of prematurely-born infants or infants with low birth weight \cite{Fiersone20183061}.
ROP can lead to permanent visual impairment and is one of the leading causes of infant blindness globally.  
Higher neonatal survival rates  have significantly increased the number of premature infants; consequently, there are sharp increases in ROP cases for infants. It is  estimated that nineteen million children are visually impaired worldwide \cite{Blencowe2013}, among which ROP accounts for six to eighteen percent in  childhood blindness \cite{gilbert_rahi_eckstein_osullivan_foster_1997}. Early treatment has  confirmed the efficacy of treatment for ROP \cite{10.1001/archopht.121.12.1684}. Therefore, it is crucial that at-risk infants receive timely retinal examinations and early detection of potential ROP. 

Early detection of ROP faces   significant  challenges. The most imposing one  is the dire lack of experienced ophthalmologists  for the screen of ROP, even in developed countries. 
This challenge is compounded by the limited awareness of ROP even among medical professionals and infants' inability of active participation in medical diagnosis.

Deep learning has made significant progresses in image classification and pattern recognition, and has shown 
great potentials for applications in medical images, such as of diabetic retinopathy (DR) \cite{10.1001/jama.2016.17216}. 
However, the development of  high-performance deep learning methodology for medical imaging has two critical requirements. First, it requires  collections of large datasets with tens of thousands of abnormal (positive) cases. Secondly,  clinical validation datasets for evaluation of  final performance require multiple grades for each image to ensure the consistency of the grading result. 
 For instance, DR is  a well-recognized complication among tens of millions of diabetic patients and the available datasets with both positive and negative samples from DR diagnosis are  in the order of several hundred thousands. 
 In the case of ROP, however, it is infeasible to meet these two key requirements. Due to the limited awareness and expertise  among medical professionals,  ROP dataset is limited and  extremely imbalanced in terms of the ratio between negative and positive images. (See Section \ref{sec:data}).
   More importantly, clinical screening for ROP often requires unusually high sensitivity level, i.e., higher than the medical standard of 95\%. This is  to ensure that few positive ROP cases  are missed for infants.

\paragraph{Our work.} 
We formulate  the problem of identifying ROP from retinal fundus images in an optimization framework, and  adopt  neural network techniques to solve this optimization problem.

Our study consists of two stages.
First, we use a shallow convolutional neural network called ROPBaseCNN. This ROPBaseCNN-based model works well and achieves  over 91 percent in both specificity and sensitivity on the first dataset Data\_0 collected from a single data source.
To increase the robustness of the model, more  ROP data  Data\_1 from multiple data sources are collected, and a deeper neural network ROPResCNN is developed. This updated ROPResCNN-based model overcomes the over-fitting and vanishing/exploding gradient problem.

Our experiments demonstrate that ROPResCNN-based model dominates both human
experts and ROPBaseCNN-model, by a wide margin.
It shows impressive performance on the combined 
Data\_0 and Data\_1:  a perfect  score on sensitivity, excellent scores of specificity (96\%) and precision (96\%), and  across-the-board improvement of roughly 10\% when compared with experienced
ophthalmologists. 
Most importantly, it reduces human errors by over 66\% in all categories, and in particular eliminates completely the error in the category of sensitivity, the most critical requirement for diagnosis of ROP.

In addition to excellent  experimental results,  our study shows that as the network gets deeper, significant features can be  extracted for better understanding of ROP. For instance, in spite of the limited and imbalanced data, ROPResCNN-based model succeeds in learning and capturing explicitly a well-known indicator for the medical diagnosis of ROP.

\section{ROP: Data Collection, Augmentation, and Processing}
\label{sec:data}

\begin{figure}[h]
    \centering
    \begin{subfigure}{0.5\textwidth}
    \centering
    \includegraphics[width=0.7\textwidth]{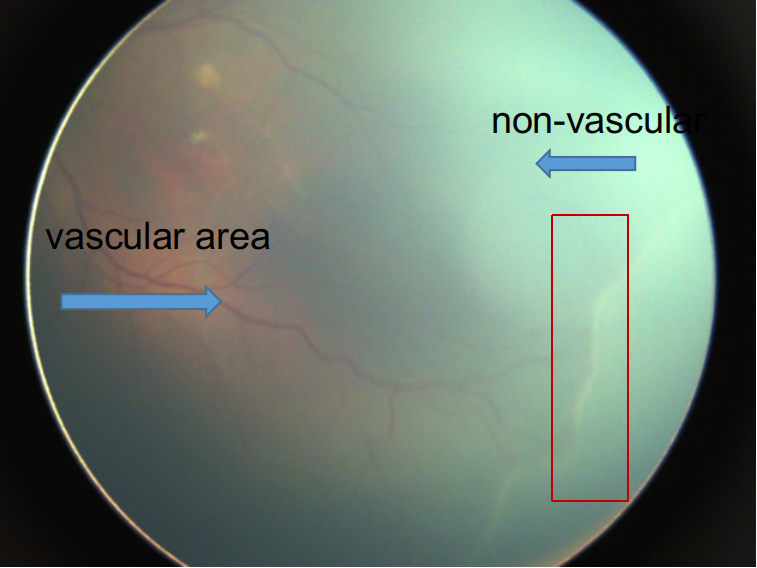}
    \caption{Positive sample}
    \end{subfigure}%
    \hspace*{\fill}
    \begin{subfigure}{0.5\textwidth}
    \centering
    \includegraphics[width=0.7\textwidth]{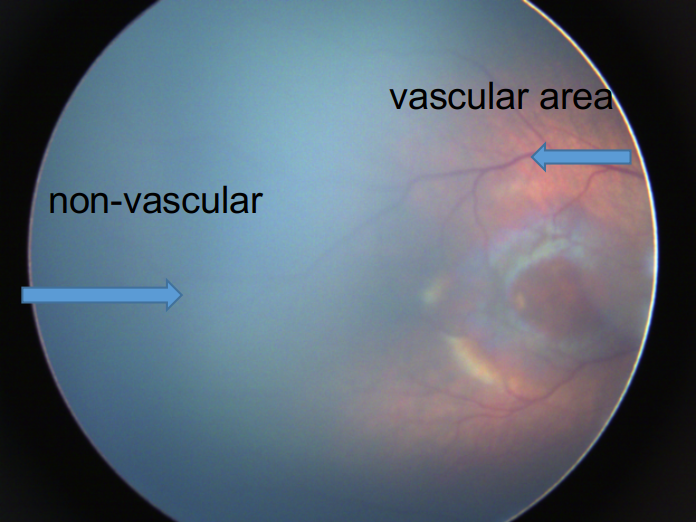}
    \caption{Negative sample}
    \end{subfigure}%
    \caption{positive sample and negative sample in Data\_0; the difference between the vascular area and the non-vascular area is clear; note the apparent thickened ridge
    (indicated in the red box) in the positive sample between the vascular  and the non-vascular areas, with no such appearance in the negative sample.}
    \label{fig:data_0}
\end{figure}

\begin{figure}[h]
    \centering
    \begin{subfigure}{0.5\textwidth}
    \centering
    \includegraphics[width=0.6\textwidth]{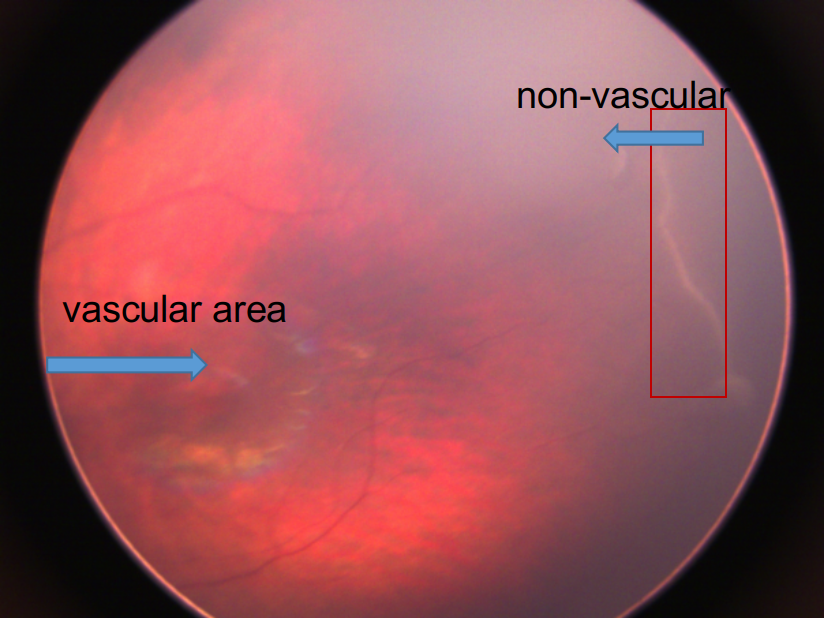}
    \caption{Positive sample}
    \end{subfigure}%
    \hspace*{\fill}
    \begin{subfigure}{0.5\textwidth}
    \centering
    \includegraphics[width=0.6\textwidth]{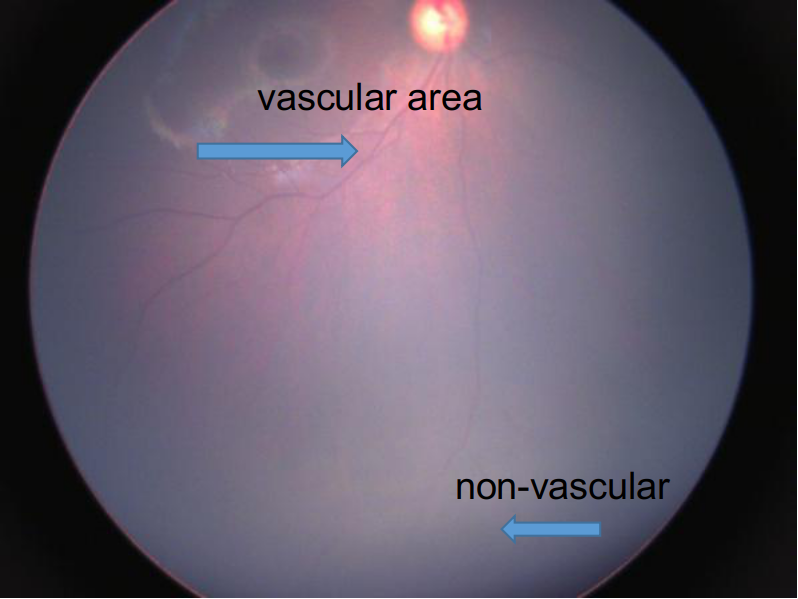}
    \caption{Negative sample}
    \end{subfigure}%
    \caption{positive sample and negative sample in Data\_1; note the thickened white line (indicated in the red box) in the positive sample between the vascular and the non-vascular areas.}
    \label{fig:data_1}
\end{figure}

\subsection{Data Collection}
  
To develop a model for ROP detection, ROP retinal fundus images were retrospectively collected from the Affiliated Eye Hospital of Nanchang University, which is an AAA
(i.e., the highest ranked) hospital in China.
All images were de-identified according to patient privacy protection policy, and ethics review was approved by the ethical committee of the university. 

Two datasets were used. The first de-identified dataset, Data\_0,  consists of  random samples of ROP images taken at the hospital between 2013 and 2018. A single type of fundus camera, Clarity Retcam3, was used with 130\degree fields of view. All operators had gone through professional training. Data\_0  includes 2021 negative samples and 382 positive samples with the resolution of $1600 \times 1200$ retinal fundus image. Images in this dataset share a common characteristic:  the boundary between the vascular and the  non-vascular areas is clear and the color difference is obvious. As shown in Figure \ref{fig:data_0}, there is a clear white dividing line, called the demarcation line, between the vascular and the non-vascular areas of the peripheral retina. In the early stage of ROP, this demarcation line will get thicker until a ridge occurs. As the ridge gets thicker,  proliferation of abnormal blood vessels will cause the retinal blood vessel to expand, eventually leading to the ROP problem. The appearance of thickened ridges is the main indicator used by 
ophthalmologists to diagnose ROP.
Note that there is no such thickened ridge in the negative sample.

The second de-identified dataset  Data\_1 consists of 461 negative samples and 498 positive samples with the resolution of $1600\times1200$ retinal fundus images. A variety of  130\degree fields cameras were used, including CLARITY Retcam3, SUOER SW-8000, and MEDSO ORTHOCONE RS-B002. This set of  data is characterized by the similar appearance of the  vascular and non-vascular areas, and with similar colors. However, the boundary between the vascular area  and the non-vascular area is much clearer than that in Data\_0. Figure \ref{fig:data_1} shows a negative sample and a positive sample from Data\_1.
 
All images 
were graded by ophthalmologists for the presence of ROP severity and for the image quality using an annotation tool. The annotation tool was designed by ophthalmologists and implemented by ourselves. ROP severity was graded as  positive or negative.  Image quality was assessed by graders, with images of adequate quality considered gradable. The reliability of the grading result was assessed by four prominent ophthalmologists. The final grading results, for which the diagnosis  from the hospital agreed with the majority of diagnosis from these ophthalmologists, were used for each retinal fundus image.


\subsection{Data Processing and Balancing}
\label{sec:databalance}
The datasets are imbalanced, for instance,  negative samples in Data\_0 dataset is five times more than  positive samples.
Consequently, the training process may be 
 significantly biased  towards the class with more samples. Data imbalance is very common among medical data. There are several approaches, including   under-sampling \cite{Guo2016}, re-sampling and fine-tuning \cite{HAVAEI201718}, oversampling \cite{10.5555/3000292.3000304}, and weight balance and class balance \cite{1549828}.

\begin{figure}
    \centering
    \begin{minipage}{0.5\textwidth}
     \centering
     \includegraphics[width=0.6\textwidth]{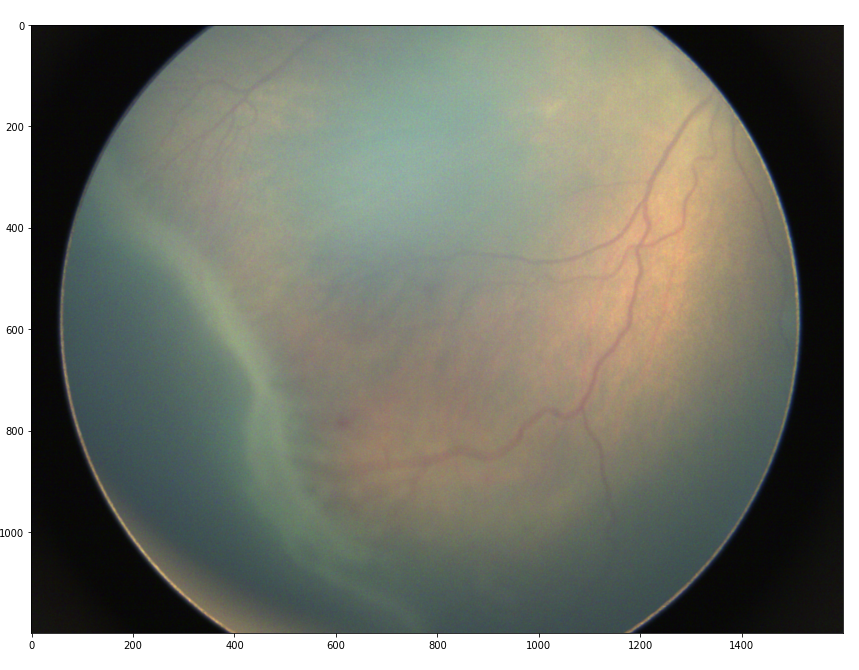}
    \end{minipage}%
    \begin{minipage}{0.5\textwidth}
     \centering
     \includegraphics[width=0.6\textwidth]{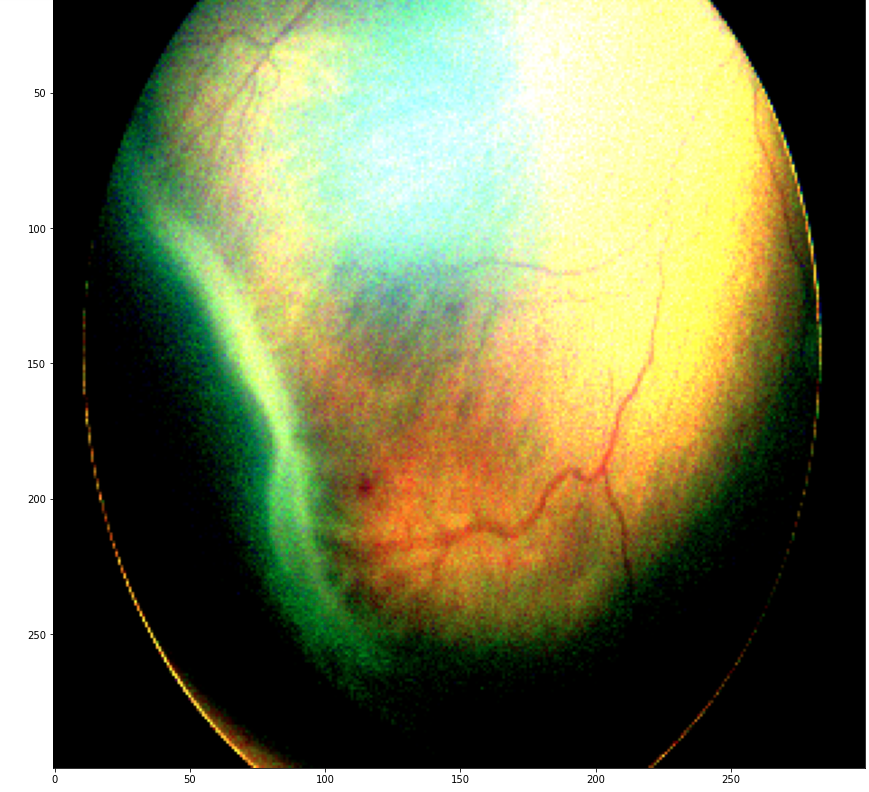}
    \end{minipage}
    \caption{the left is an original retinal fundus image; the right is the same image from the left after data processing.}
    \label{fig:my_label}
\end{figure}

To mitigate the imbalance problem, we design a hybrid method with a combination of several techniques:
\begin{enumerate}[(i)]
    \item Data enhancement:
    all samples in the dataset are first enhanced by brightness adjustment and random flipping.
    (See Figure \ref{fig:my_label}).
    Afterwards, all images are resized into $300\times 300$.
    
    \item Tuning sampling ratio and class weights:
    we use different class weights in the cross entropy loss function in our optimization framework, to be introduced in the next section. We over-sample the enhanced positive samples and re-sample the enhanced negative samples, so that the numbers of positive samples and negative samples in the sample batch are kept proportional to the inverse of their class weights. We experiment with different ratios through grid search in the validation set, and eventually set the ratio of positive and negative samples to $1:2$ in the training process. 
    \end{enumerate}
This data processing and balance strategy is used throughout our study. 

\section{Problem and Optimization}
\paragraph{Problem formulation.}
We formulate the problem of detecting ROP as a binary classification problem, where the positive images and the negative images are labeled as $1$ and $0$, respectively.
That is, given a fundus image, instead of labelling the image as either $1$ or $0$, we assign a score in terms of probability between $0$ and $1$ to the input image. The higher the score, the higher the probability that the image has an ROP (i.e., ROP positive). 
When assigning the label for the input image, if the probability is higher than $0.5$, it is then labelled as  positive; otherwise it is negative.
This is a natural choice for the neural network which requires the output to be a continuous variable.

Now, suppose the probability  is parametrized by $\theta$ such that it is denoted as $p_\theta$.
This set of parameter could be interpreted as various factors contributing to the probability of having an ROP.  Then
the training stage is to minimize  the cross entropy loss function over the set of parameters $\theta$. That is,  denote the distribution of the pair of the image $x$ and the 0-1 label $y$ by $p_{\text{data}}$, 
then the training process is to solve the following optimization problem,
\begin{align*}
    \min_{\theta}\mathbb{E}_{(x, y)\sim p_{\text{data}}}[-y\log p_{\theta}(x)-(1-y)\log(1-p_{\theta}(x))].
\end{align*}

Given the limited amount of available data and hence possible issue of overfitting, we add a kernel regularization. In particular, we adopt the $L^2$ regularization on the weight matrices of the fully connected layers. For each fully connected layer with weight matrix $W=(w_{ij})$, we add the following regularization term to the loss function
\[
    \lambda \|W\|_{L^2} = \lambda\sum_{i,j}w_{ij}^2,
\]
where $\lambda$ is a hyperparameter to adjust the scale of the regularization.
Finally, we adjust the loss function by the $1:2$ class weight from the data processing stage, so that the final optimization problem is to solve the following regularized cross entropy loss function,
\begin{align}\label{eqn:optimization}
    \min_{\theta}\mathbb{E}_{(x, y)\sim p_{\text{data}}}[-y\log p_{\theta}(x)-2(1-y)\log(1-p_{\theta}(x))]+\lambda\|W\|_{L^2}.
\end{align}

\paragraph{Optimization.}
We used the Adam algorithm \cite{adam} to solve  our optimization problem \eqref{eqn:optimization}.
Adam algorithm combines a momentum method and an adaptive learning rate method, and uses the first order and the second order moments to optimize the neural network which is to be specified in details in section \ref{sec:CNN}.
Adam algorithm is more efficient than the vanilla stochastic gradient descent algorithm.

The parameters of Adam algorithm used here are: learning rate $\alpha = 0.001$, and the exponential decay rate for the first and the second moment $\beta_1=0.9$ and $\beta_2 = 0.999$, respectively.

To train our models more efficiently, we adjust the learning rate with respect to the validation loss. 
 More specifically, the learning rate is reduced by 20\% when the validation loss does not improve for $5$ epochs. 

\begin{algorithm}  
  \caption{Adam is proposed for stochastic optimization. Default settings are $\alpha$ = 0.001, $\beta_1$ = 0.9, $\beta_2$ = 0.999, $\eta=10^{-7}$, $r=0.8$, and $n=5$.}
  \begin{algorithmic}[] 
  \Require  $\alpha$: Learning rate
    \Require $\beta_1, \beta_2$: $\in$ [0, 1): Exponential decay rates for the moment estimates 
     \Require $f(\theta)$: Stochastic objective function with parameters $\theta$  
    \Require $\theta$: Initial parameter vector    
    \Require $r$: Learning rate reduction factor
    \Require $n$: Patience parameter
    \Require $m_0$ $\gets$ 0: Initialize $1^{st}$ moment vector
    \Require $v_0$ $\gets$ 0: Initialize $2^{nd}$ moment vector
    \Require $i$ $\gets$ 0: Initialize step
     \While{$\theta_i$ not converged}     \State $i\gets i+1$
    \State $g_i$ $\gets$ $\nabla_\theta$$f_i(\theta_{i-1})$ (Get gradients  stochastic objective at step i)
    \State $m_i$ $\gets$ $\beta_1$ $\cdot$ $m_{i-1}$+($1-\beta_1$)$\cdot g_i$ (Update biased first moment estimate)
    \State $v_i$ $\gets$ $\beta_2$ $\cdot$ $v_{i-1}$+($1-\beta_2$)$\cdot g_i^2$ ((Update biased second raw moment estimate)
    \State $\hat{m}_i$ $\gets$ $m_i/(1-\beta_1^i$)(Compute bias-corrected first moment estimate)
     \State $\hat{v}_i$ $\gets$ $v_i/(1-\beta_2^i$)(Compute bias-corrected second raw moment estimate)
     \State $\theta_i$ $\gets$ $\theta_{i-1}-\alpha \cdot$ $\hat{m}_i / (\sqrt[2]{\hat{v}_i}+\eta)$ (Update parameters)
     \State $\alpha\leftarrow r\alpha$ if the validation result does not improve for $n$ steps.
     \EndWhile\\ 
    \Return{$\theta_i$ (Resulting parameter)} 
    \end{algorithmic}  
 \end{algorithm}
 
\section{Convolutional Neural Network and Architectures}
\label{sec:CNN}
The network adopted here is the convolutional neural network (CNN) \cite{LeCun1999ObjectRW}.
CNN is specialized for processing grid-like data including images.
A CNN  consists of the feature extractor and the decoder.
The feature extractor has several convolutional layers and pooling layers.
It captures the basic features such as lines and corners in the first few layers and extracts more advanced features such as the indicators of ROP in the later layers.
The extracted features are then fed into the decoder part.
The decoder is a set of fully connected layers.
The decoder uses the extracted features to predict the target variable.
In our problem, the probability of an input image having ROP is the target variable. We experiment with two different architectures for CNN: 
ROPBaseCNN and ROPResCNN.

\begin{figure}[h]
    \centering
    \begin{tabular}{|c|c|}\hline
    Type of layer & parameters \\ \hline \hline
    Input& shape=(300,300,3)  \\ \hline
    Convolution & filters=32, kernel size=$3\times 3$, stride2=(2,2), activation=ReLU \\ \hline
    Max pooling & pool size=$2\times 2$, strides=(2,2)\\ \hline
    Convolution & filters=64, kernel size=$3\times 3$, stride2=(2,2), activation=ReLU \\ \hline
    Max pooling & pool size=$2\times 2$, strides=(2,2)\\ \hline
    Dropout & dropping probability = 0.25\\ \hline
    Flatten & none \\ \hline
    Fully connected & neurons=128, activation=ReLU, $\lambda=0.001$\\ \hline
    Dropout & dropping probability = 0.5\\ \hline
    Fully connected & neurons=64, activation=ReLU, $\lambda=0.001$\\ \hline
    Output(Dense) & shape=(1), activation=sigmoid\\ \hline
    \end{tabular}
    \caption{the architecture of ROPBaseCNN}
    \label{figure:ROPBaseCNN}
\end{figure}


\begin{figure}[h]
    \centering
    \begin{tabular}{|c|c|c|c|}\hline
    model & ROPBaseCNN & ROPBaseCNN & ROPResCNN \\ \hline
    Train data & Data\_0 & Data\_0+Data\_1 & Data\_0+Data\_1 \\ \hline
    Test data & Data\_0 & Data\_0+Data\_1 & Data\_0+Data\_1 \\ \hline
    Precision & 0.9479 & 0.8131 & 0.96 \\ \hline
    Sensitivity & 0.91 & 0.7891 & 1.0 \\ \hline
    Specificity & 0.9135 & 0.9335 & 0.96 \\ \hline
    Accuracy & 0.93 & 0.8948 & 0.98 \\ \hline
    F1 score & 0.9286 & 0.8009 & 0.98 \\ \hline
    \end{tabular}
    \caption{experimental results with data Data\_0 and Data\_1}
    \label{fig:result_table}
\end{figure}

\begin{figure}[h]
    \centering
    \includegraphics[width=0.7\textwidth]{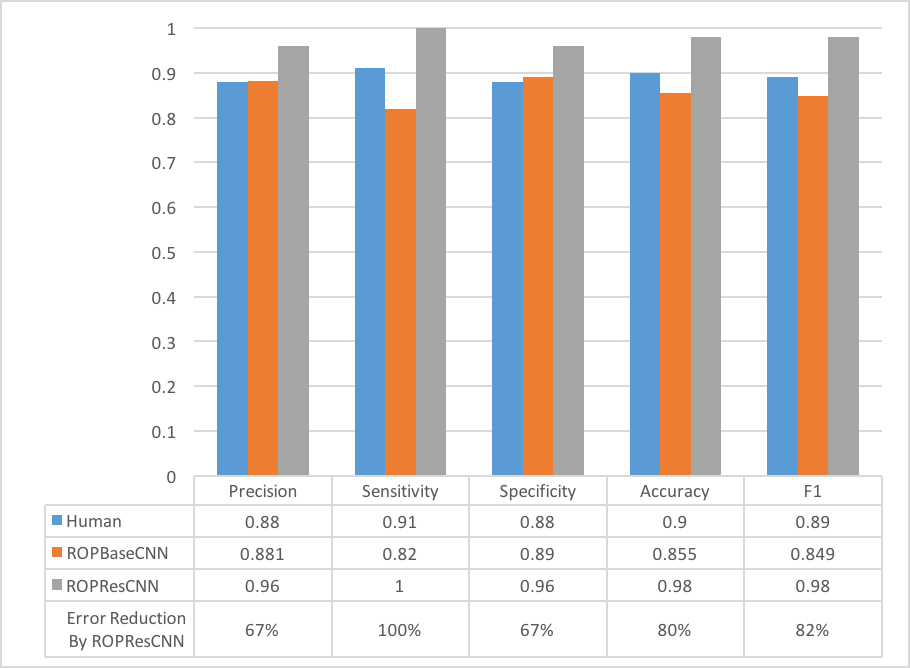}
    \caption{comparison between ophthalmologists and our models} \label{fig:comparision}
\end{figure}


\begin{figure}[h]
    \centering
    \begin{minipage}{0.3\textwidth}
     \centering
     \includegraphics[width=0.7\textwidth]{figures/processed.png}
    \end{minipage}%
    \begin{minipage}{0.3\textwidth}
     \centering
     \includegraphics[width=0.7\textwidth]{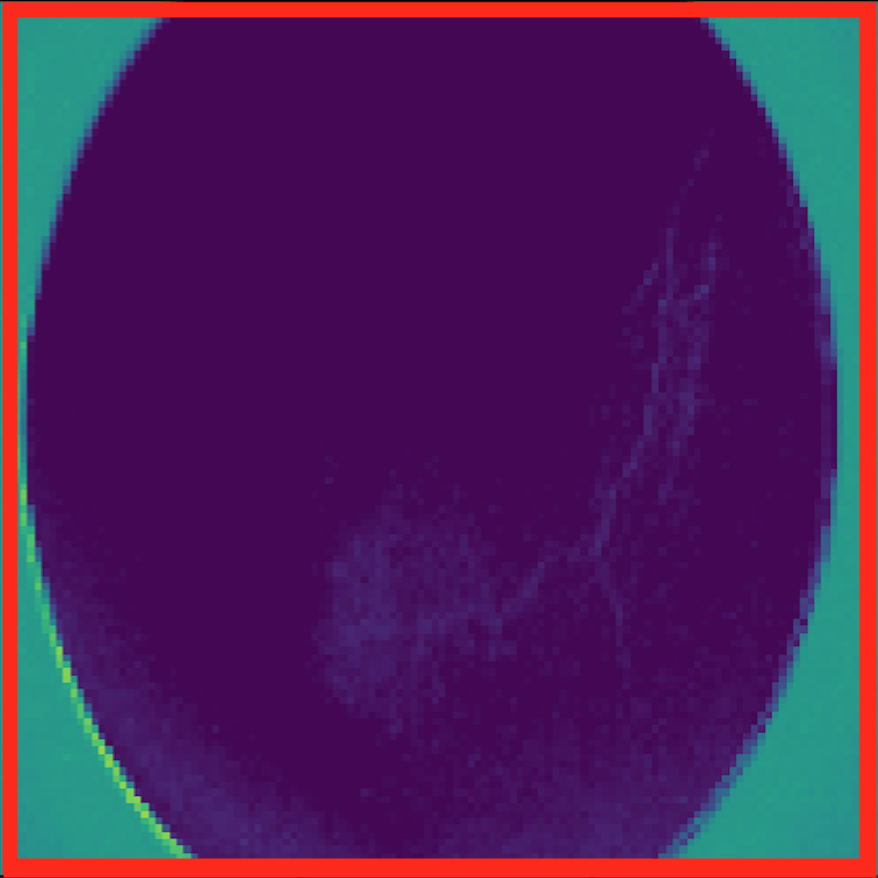}
    \end{minipage}
    \begin{minipage}{1\textwidth}
     \centering
     \includegraphics[width=\textwidth]{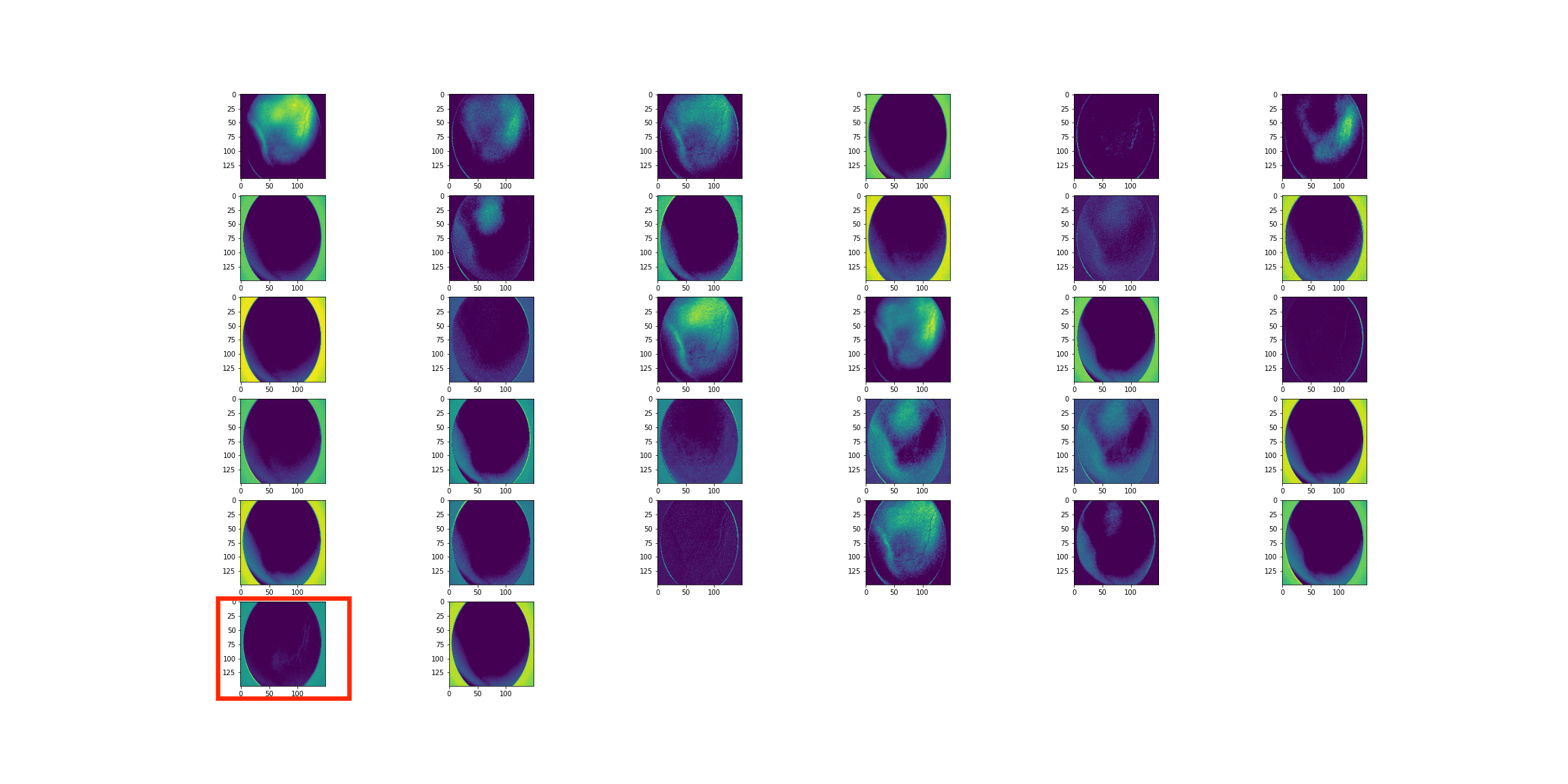}
    \end{minipage}%
    \caption{the top left is the preprocessed $300\times 300$ image fed into the ROPBaseCNN; the top right is the extracted feature that shows abnormal blood vessel growth; the bottom is the output from the second layer of ROPBaseCNN.}
    \label{figure:base_hidden_output}
\end{figure}

\begin{figure}[h]
    \centering
    \begin{minipage}{0.3\textwidth}
     \centering
     \includegraphics[width=0.6\textwidth]{figures/processed.png}
    \end{minipage}%
    \begin{minipage}{0.3\textwidth}
     \centering
     \includegraphics[width=0.6\textwidth]{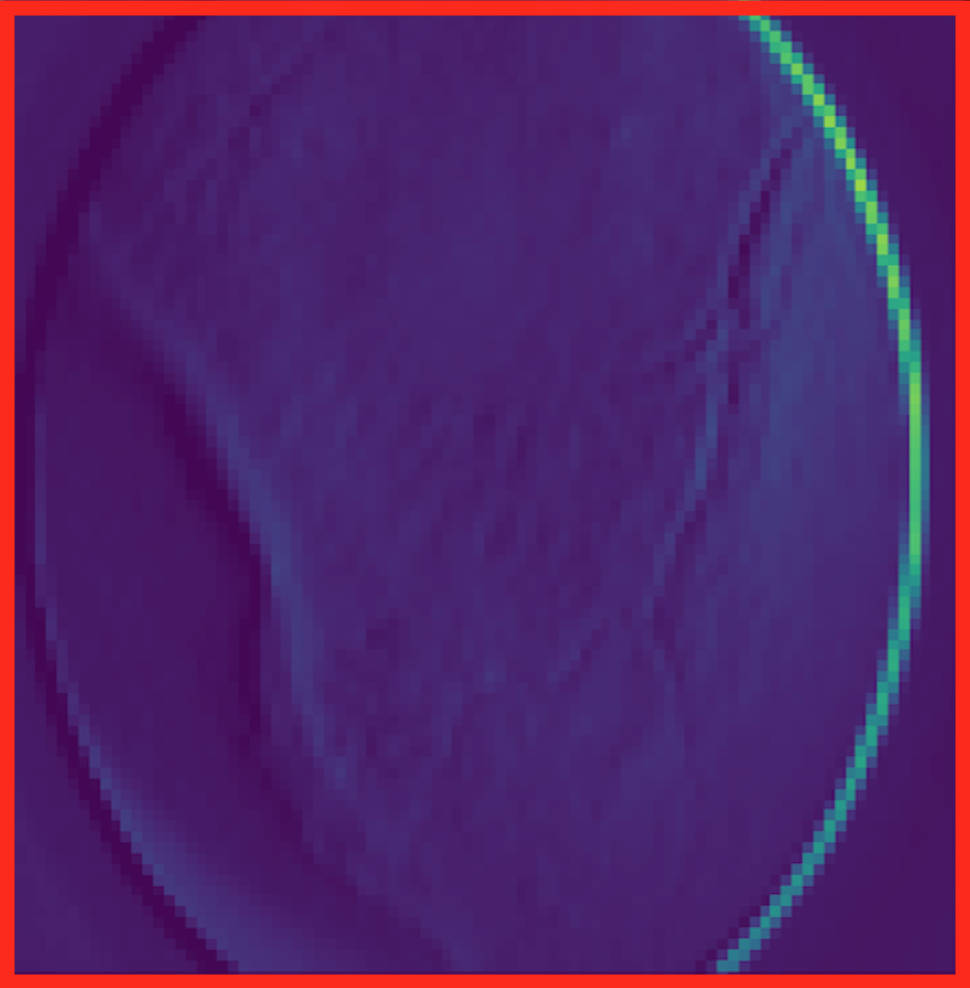}
    \end{minipage}
    \begin{minipage}{0.3\textwidth}
     \centering
     \includegraphics[width=0.6\textwidth]{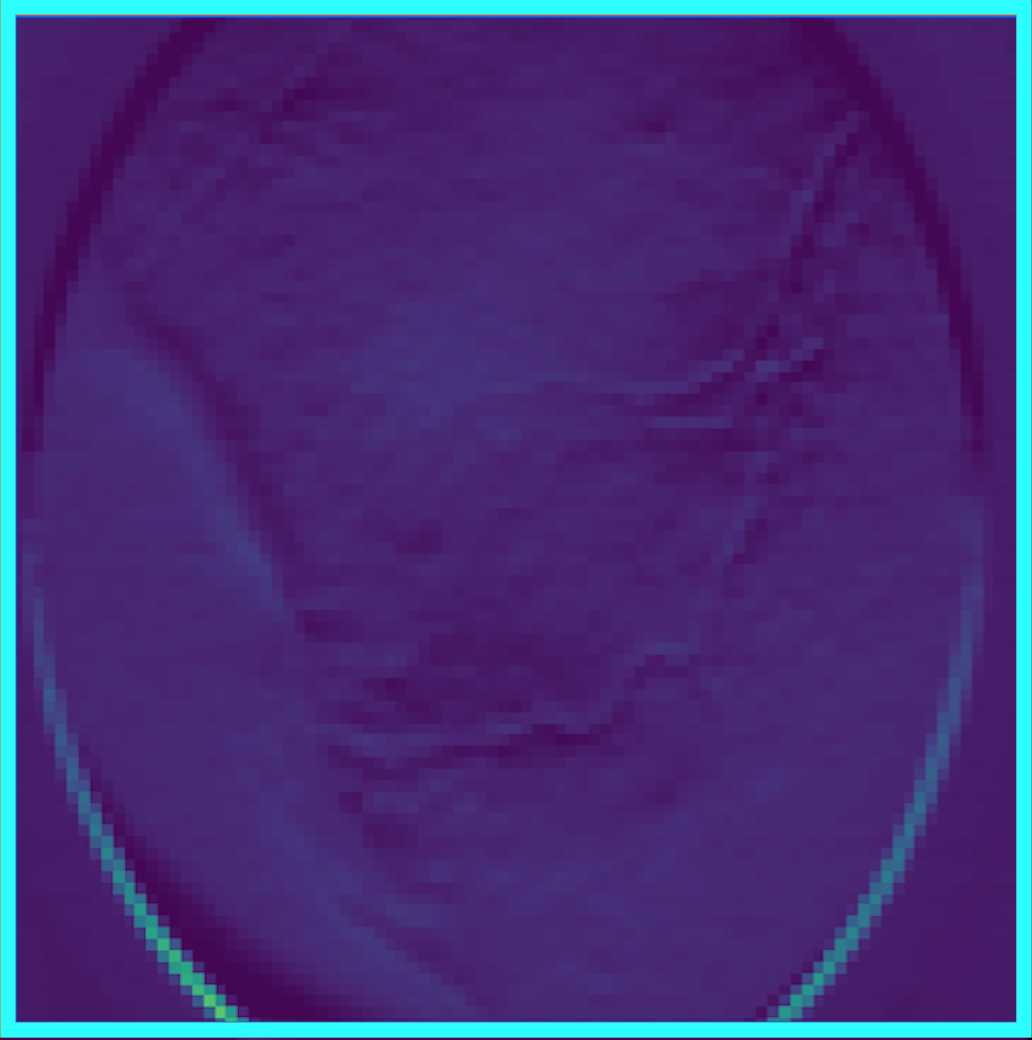}
    \end{minipage}
    \begin{minipage}{1\textwidth}
     \centering
     \includegraphics[width=\textwidth]{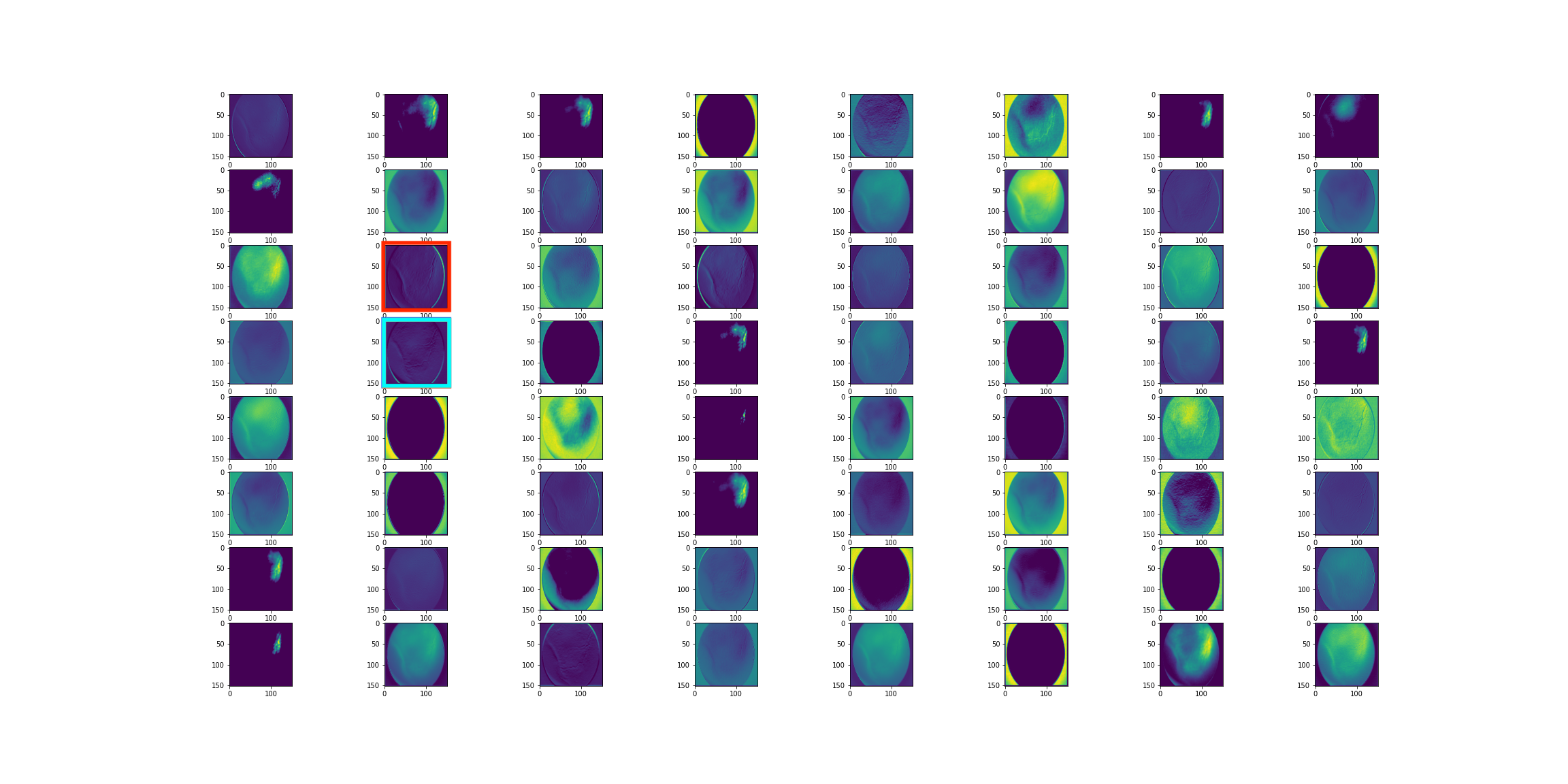}
    \end{minipage}%
    \caption{the top left is the preprocessed $300\times 300$ image fed into the ROPResCNN; the top middle and the top right are the extracted features showing the occurrence of the thickened ridge; the bottom is the output from the fifth layer of ROPResCNN.}
    \label{figure:res_hidden_output}
\end{figure}

\paragraph{The architecture of ROPBaseCNN.}
For dataset Data\_0 with only 2401 samples, complicated models are prone to the overfitting problem on the training data, therefore we first adopt this shallow CNN model with only five layers: two convolution layers and three fully-connected layers. 
To prevent the issue of  overfitting, we add dropout layers \cite{JMLR:v15:srivastava14a} in the decoder part and the $L^2$ regularization  for the kernel of the fully connected layers.
The architecture of this shallow CNN, named ROPBaseCNN, is summarized in Figure \ref{figure:ROPBaseCNN}.

Combined with aforementioned data processing strategy, the accuracy of ROP detection under ROPBaseCNN for dataset Data\_0 is 93\%. This model appears unstable in that its performance would deteriorate with the combined datasets Data\_0 and Data\_1. (See table \ref{fig:result_table}).
\paragraph{The architecture of ROPResCNN.}
The architecture of ROPResCNN combines a pre-trained ResNet50 \cite{7780459} model (no top) with a global average pool layer and a fully-connected layer as the output layer. 
The weights of ResNet50 are used as the initial point of the optimization.

With the limited amount of ROP data, training a deep neural network from a random initialization is difficult. Instead, we adopt the pre-trained network weights, which help accelerate the training process because ResNet50 is capable of capturing basic and important features for general image classification. Moreover, with this  pre-trained network, we manage to avoid the well-known issues in deep networks such as  vanishing/exploding gradient.
Additionally, we use the global average pooling at the end of the pre-trained residual network, reducing the dimension from 3D to 1D. Therefore, global pooling outputs one response for every feature map.
At the end, a dense layer with one neuron with the Sigmoid activation function aggregates and outputs the probability.

 Note that ROPResCNN favors more convolutional layers instead of fully connected ones. With the  global-average-pooling for dimension reduction, empirically it is not necessary to add regularization for ROPResCNN.

\section{Implementation Results}
Data\_0 is used in ROPBaseCNN
and  split into three sets: the training set has 187 positive samples and 990 negative samples, the validation set has 80 positive samples and  425 negative samples, and the testing set has 115 positive samples and 606 negative samples with held-out class labels. A combination of  Data\_0 and Data\_1 is used for ROPResCNN, and is split into three sets: training (431 positive samples and 1216 negative samples), validation (185 positive samples and  521 negative samples), and testing (264 positive samples and 745 negative samples with held-out class labels). 
\paragraph{Evaluation metrics.}

We use the following standard metrics to evaluate the performance of our models, including precision, sensitivity, specificity, accuracy, and the F1 score,
\begin{align*}
    \text{Precision} &= \frac{TP}{TP+FP},  \ \ \ \
    \text{Sensitivity} = \frac{TP}{TP+FN},\\
    \text{Specificity} &= \frac{TN}{TN+FP}, \ \ \ \ 
    \text{Accuracy} = \frac{TP+TN}{TP+TN+FP+FN},\\
    \text{F1} &= 2\frac{\text{Precision}\times\text{Sensitivity}}{\text{Precision}+\text{Sensitivity}},
\end{align*}
where TP, TN, FP, and FN represent true positive, true negative, false positive, and false negative, respectively. The error reduction of the models is calculated as $$\text{Error reduction}=\frac{\text{human error}-\text{model error}}{\text{human error}}.$$

\paragraph{Training on single GPU.}
A single GTX 1080 GPU and 8GB of memory is used for training the ROPBaseCNN-based model and the ROPResCNN-based model. With appropriate data processing, one GPU turns out to be sufficient to fit the training process with 3000 samples per epoch. For ROPBaseCNN, the batch size is 32, and the training is stopped after 25 epochs; for ROPResCNN, the batch size is 64, and the training is stopped after 30 epochs.

\paragraph{Evaluations.}
The results of ROPBaseCNN-based model with Data\_0 are summarized in the first column of Table \ref{fig:result_table}, the results of ROPBaseCNN-based model with the combined datasets Data\_0 and Data\_1 are summarized in the second column of Table \ref{fig:result_table}; and the third column shows the results of ROPResCNN-based model with both datasets Data\_0 and Data\_1.

We take 200 infants' retinal fundus images with confirmed  grading results by ophthalmologists. Their results are then compared against those generated by our models. Figure \ref{fig:comparision} gives the detailed performance comparison.
We see that ROPBaseCNN-based model  manages to achieve comparable performance with experienced ophthalmologists, especially  in terms of precision and specificity. However, its performance
is not robust, with excellent scores from Data\_0 vanishing on the combined  Data\_0 and Data\_1. 

ROPResCNN-based model dominates both human
experts and ROPBaseCNN-model, by a wide margin.
It shows impressive performance on the combined 
Data\_0 and Data\_1:  a perfect  score on sensitivity, excellent scores in specificity (96\%) and  precision (96\%), and  across-the-board improvement of roughly 10\% when compared with experienced
ophthalmologists. 
Most importantly, it reduces human errors by over 66\% in all categories, and in particular eliminates completely the error in the category of sensitivity, the most critical requirement for diagnosis of ROP.

\paragraph{Feature map.}
 The feature map from ROPBaseCNN, shown in Figure \ref{figure:base_hidden_output},  captures an implicit indicator of ROP, the abnormal blood vessel growth. However, such a disorder from the retinal fundus image is not used by ophthalmologists as a standard indicator for diagnosis of ROP. 

The feature map from ROPResCNN demonstrates that  ROPResCNN-based model succeeds in learning and capturing  explicitly  the well-accepted indicator for the medical diagnosis of ROP: the thickened ridge. (See Figure 
\ref{figure:res_hidden_output}).

\section{Summary}

Our study shows that models using the state-of-art CNN for general image classification can provide accurate and early detection of ROP with a perfect sensitivity score and excellent scores in specificity and precision.
Beyond diagnosis, our study shows that deep neural network techniques  can  be potentially powerful to extract significant  features for better understanding of ROP.

\newpage

\section*{Broader impact}
As far as the authors are concerned, a) researchers at the intersection of deep learning and medical imaging can potentially benefit from this work; b) no particular group of people in the society is expected to be put in disadvantage due to this work; c) this work is not subject to the failure of  the system; d) data bias is out of the scope of the potential influence of this work; and e) proper understanding of models helps reducing the risk of misdiagnosis.
\bibliographystyle{plain}
\bibliography{main}
\end{document}